\documentstyle[12pt]{article}
\newlength{\extraspace}
\newlength{\extraspaces}
\catcode`\@=11
\def\numberbysection{\@addtoreset{equation}{section}
\def\theequation{\arabic{section}.\arabic{equation}}}
\newcommand{\newsection}[1]{
\vspace{7mm}
\pagebreak[3]
\addtocounter{section}{1}
\setcounter{equation}{0}
\setcounter{subsection}{0}
\setcounter{footnote}{0}
\begin{center}
{\large {\bf \thesection. #1}}
\end{center}
\nopagebreak
\medskip
\nopagebreak
\hspace{3mm}}
\newcommand{\nonu}{\nonumber \\[.5mm]}
\newcommand{\A}{&\!\!\!}

\newcommand{\br}{\begin{array}}
\newcommand{\er}{\end{array}}
\newcommand{\mn}{\mu\nu}
\begin{document}
\addtolength{\baselineskip}{.7mm}
\thispagestyle{empty}
\begin{flushright}
STUPP--01--163 \\
{\tt hep-th/0110222} \\ 
October, 2001
\end{flushright}
\vspace{7mm}
\begin{center}
{\Large{\bf Stability of Antisymmetric Tensor Fields \\[2mm]
of Chern-Simons Type in AdS Spacetime 
}} \\[20mm] 
{\sc Hiroki Hata}\footnote{
\tt e-mail: hata@krishna.th.phy.saitama-u.ac.jp} 
\hspace{1mm} and \hspace{2mm}
{\sc Yoshiaki Tanii}\footnote{
\tt e-mail: tanii@post.saitama-u.ac.jp} \\[7mm]
{\it Physics Department, Faculty of Science \\
Saitama University, Saitama 338-8570, Japan} \\[20mm]
{\bf Abstract}\\[7mm]
{\parbox{13cm}{\hspace{5mm}
Stability of massive antisymmetric tensor fields with the 
Chern-Simons type action in anti de Sitter spacetime is 
studied. It is found that there exists a complete set of 
solutions whose energy is conserved and positive definite 
if the mass is positive. Scalar products of the solutions 
are shown to be well-defined and conserved. 
In contrast to the previously studied scalar field case 
there is no other set of stable solutions with a different 
kind of boundary condition. 
}}
\end{center}
\vfill
\newpage
\setcounter{section}{0}
\setcounter{equation}{0}
\numberbysection
%
%
\newsection{Introduction}
Anti de Sitter (AdS) spacetime is a maximally symmetric 
spacetime with a negative constant curvature. 
It naturally appears as a solution of Einstein equation with 
a negative cosmological constant. 
It also appears in compactifications of higher dimensional 
supergravities in the Kaluza-Klein theory \cite{DNP}. 
Recently interest in field theories in AdS spacetime has been 
much increased due to their relevance to the AdS/CFT 
correspondence in the string/M theory \cite{MAL,GKP,WITTEN} 
(For a review see ref.\ \cite{AGMOO}.). 
An important issue of field theories in AdS spacetime is 
their stability, which was previously discussed in 
refs.\ \cite{AD,BF,MT}. (For recent studies on the stability 
see ref.\ \cite{DFGHM}.) Another important issue is a choice 
of boundary conditions of fields at spatial infinity. 
In ref.\ \cite{AIS} boundary conditions are chosen such that 
the Cauchy problem for field equations is well-defined 
by requiring the conservation of the scalar product of fields. 
\par
The purpose of this paper is to study the stability of free 
massive antisymmetric tensor fields of arbitrary rank $n$ 
in AdS spacetime. The string/M theory and supergravities 
contain antisymmetric tensor fields, which play an important 
role. Therefore, their stability is an important issue. 
There are two types of theories of antisymmetric tensor fields. 
One type of theories have an action with the second order 
kinetic term of the Maxwell type and the other type of theories 
have an action with the first order kinetic term of the 
Chern-Simons type \cite{TPvN}. (Equivalent theories to the 
latter were studied in refs.\ \cite{DJT,DJ}.) 
Both types of theories appear in supergravities \cite{SS}. 
In this paper we only consider the Chern-Simons type theories, 
which are theories of $n$-th rank antisymmetric tensor fields 
in $d=2n+1$ dimensions. 
As a preparation for study of the stability we first obtain 
the general solution of the field equation. 
Then we obtain conditions for the stability by studying 
the conservation and the positivity of the energy. 
We also study scalar products of the solutions. 
\par
The conditions for the stability of scalar fields in AdS 
spacetime were obtained in refs.\ \cite{BF,MT}. 
A free massive scalar field theory in $d$-dimensional AdS 
spacetime is stable if the mass $m$ satisfies 
\begin{equation}
\left( {m \over a}\right)^2 > -\biggl({d-1 \over 2}\biggr)^2, 
\end{equation}
where $a^{-1}$ is the radius of AdS spacetime. 
More precisely, there exists a complete set of solutions of the 
field equation whose energy is conserved and positive definite. 
Furthermore, when the mass satisfies 
\begin{equation}
-\biggl({d-1 \over 2}\biggr)^2 
< \left( {m \over a}\right)^2 
< 1-\biggl({d-1 \over 2}\biggr)^2, 
\end{equation}
there exists another set of stable solutions satisfying a 
different kind of boundary condition at spatial infinity. 
In the latter case the coefficient of the improvement term 
in the energy-momentum tensor must take a particular value. 
Scalar fields can be stable even if the mass 
squared is negative due to a positive contribution from 
the kinetic term to the energy. 
\par
As in the scalar field case we find that there exists a complete 
set of solutions for antisymmetric tensor fields whose energy 
is conserved and positive definite if the mass is positive 
$m > 0$. There are three kinds of improvement terms in the 
energy-momentum tensor for antisymmetric tensor fields of 
rank $n \geq 2$. The coefficients of these terms can take 
arbitrary values although they do not contribute to the 
energy. The scalar products of the solutions are shown 
to be well-defined and conserved. 
\par
In contrast to the scalar field case, however, there is 
no other set of stable solutions. 
The conservation of the energy allows another set of 
solutions satisfying a different kind of boundary condition 
at spatial infinity as in the scalar field case but their 
energy turns out to be divergent. 
Therefore, only one kind of boundary condition is possible for 
antisymmetric tensor fields with the Chern-Simons type action. 
A Chern-Simons type theory of the second rank atisymmetric 
tensor fields in five-dimensional AdS spacetime was 
previously studied in ref.\ \cite{AF} in the context of 
the AdS/CFT correspondence. 
By using a different approach it was found there that 
only one kind of boundary condition is possible, 
which is consistent with our result. 
\par
In the next section we introduce the first order action of 
antisymmetric tensor fields in AdS spacetime and obtain the 
energy-momentum tensor. In sect.\ 3 the general solution of 
the field equation is obtained in terms of the hypergeometric 
functions. The conservation and the positivity of the energy are 
discussed in sects.\ 4 and 5 respectively. 
In sect.\ 6 we show that there exists a 
well-define and conserved scalar product for the solutions. 
In Appendix we give a construction and useful identities of 
spherical harmonics for antisymmetric tensors on the 
$(d-2)$-dimensional sphere ${\rm S}^{d-2}$. 
\par
\newpage
%
\newsection{Antisymmetric tensor fields in AdS spacetime}
We consider antisymmetric tensor fields in $d$-dimensional 
AdS spacetime. AdS spacetime is a maximally 
symmetric spacetime and has the metric 
\begin{equation}
g_{\mn} dx^{\mu} dx^{\nu} = {1 \over a^2 \cos^2\rho} \left[ 
- dt^2 + d\rho^2 + \sin^2\rho \ h_{ab} d\theta^a d\theta^b \right], 
\label{adsmetric}
\end{equation}
where $\mu,\nu=0,1,\cdots,d-1$ are $d$-dimensional world indices 
and the constant $a^{-1}$ is the radius of AdS spacetime. 
The time coordinate $t$ has a range $-\infty < t < \infty$, which 
corresponds to considering the universal covering of AdS spacetime. 
The radial coordinate $\rho$ has a range $0\le \rho < {\pi \over 2}$ 
with the spatial infinity at $\rho={\pi \over 2}$. 
$\theta^a$ and $h_{ab}$ ($a,b=1,2,\cdots,d-2$) are coordinates 
and the metric of the ($d-2$)-dimensional unit sphere S${}^{d-2}$. 
Non-vanishing components of the Christoffel connection are 
\begin{eqnarray}
\A\A \Gamma_{0\rho}^{\;0} = \Gamma_{00}^{\;\rho} = 
\Gamma_{\rho\rho}^{\;\rho} = \tan\rho, \nonu
\A\A \Gamma_{ab}^{\;\rho} = - \tan\rho h_{ab}, \qquad
\Gamma_{\rho b}^{\;a} 
= {1 \over \sin\rho \cos\rho} \delta_b^a, \nonu
\A\A \Gamma_{bc}^{\;a} = {1 \over 2} h^{ad} \left( 
\partial_b h_{cd} + \partial_c h_{bd} - \partial_d h_{bc} \right) 
= \gamma_{bc}^{\;a}, 
\label{cristoffel}
\end{eqnarray}
where $\gamma_{bc}^{\;a}$ is the Christoffel connection of 
S${}^{d-2}$. The Riemann tensor is given by 
\begin{equation}
R_{\mn}{}^{\tau}{}_{\sigma} 
= - a^2 \left( \delta_{\mu}^{\tau} g_{\nu\sigma} 
- \delta_{\nu}^{\tau} g_{\mu\sigma} \right). 
\label{rtensor}
\end{equation}
Our conventions for the curvature tensors are 
$R_{\mu\nu}{}^\tau{}_\sigma 
= \partial_\mu \Gamma^{\;\tau}_{\nu\sigma} 
+ \Gamma^{\;\tau}_{\mu\lambda} \Gamma^{\;\lambda}_{\nu\sigma} 
- (\mu \leftrightarrow \nu)$, 
$R_{\mu\nu} = R_{\tau\mu}{}^\tau{}_\nu$, $R 
= g^{\mu\nu} R_{\mu\nu}$. 
For other properties of AdS spacetime see, e.g. 
ref.\ \cite{AGMOO}. 
\par
%
%
We consider a free theory of a complex antisymmetric tensor 
field $B_{\mu_1 \cdots \mu_n}$ of rank $n$ in $d$-dimensional 
AdS spacetime for $d=2n+1$. The Chern-Simons type 
action \cite{TPvN} is 
\begin{eqnarray}
S \A = \A \int d^d x \biggl[ (-1)^{\frac12(n+1)} 
{1 \over (n!)^2} \epsilon^{\mu_1 \cdots \mu_{2n+1}} 
B^*_{\mu_1 \cdots \mu_n} \partial_{\mu_{n+1}} 
B_{\mu_{n+2} \cdots \mu_{2n+1}} \nonu
\A\A - {m \over n!} \sqrt{-g} B^*_{\mu_1 \cdots \mu_n} 
B^{\mu_1 \cdots \mu_n} \biggr], 
\label{action}
\end{eqnarray}
where $\epsilon^{\mu_1 \cdots \mu_{2n+1}}$ is the totally 
antisymmetric tensor and $*$ denotes the complex conjugation. 
The reality of the action requires that the mass $m$ is real. 
When $m=0$, the action consists of only the kinetic term, 
which do not depends on the metric but is invariant under 
general coordinate transformations. 
This is an action of a topological field theory. 
We do not discuss the stability of the $m=0$ case since 
there is no local degrees of freedom and the energy is zero. 
The field equation derived from this action is 
\begin{equation}
(-1)^{\frac12(n+1)} 
{1 \over n!} \epsilon^{\mu_1 \cdots \mu_{2n+1}} 
\partial_{\mu_{n+1}} B_{\mu_{n+2} \cdots \mu_{2n+1}} 
- m \sqrt{-g} B^{\mu_1 \cdots \mu_n} = 0. 
\label{fieldeq}
\end{equation}
\par
%
%
We define the energy-momentum tensor of the theory as 
a variation of the action with respect to the metric. 
To do this we need an action for general metric, which reduces 
to the original action (\ref{action}) for the AdS metric 
(\ref{adsmetric}). We use the action 
\begin{eqnarray}
S' = \int d^d x \A \A \biggl[ (-1)^{\frac12(n+1)} 
{1 \over (n!)^2} \epsilon^{\mu_1 \cdots \mu_{2n+1}} 
B^*_{\mu_1 \cdots \mu_n} \partial_{\mu_{n+1}} 
B_{\mu_{n+2} \cdots \mu_{2n+1}} \nonu
\A\A - {\mu \over n!} \sqrt{-g} 
B^*_{\mu_1 \cdots \mu_n} B^{\mu_1 \cdots \mu_n} 
+ {\alpha \over n!a} R B^*_{\mu_1 \cdots \mu_n} 
B^{\mu_1 \cdots \mu_n} \nonu
\A\A + {\beta \over n!a} R_{\mn} B^{*\mu}{}_{\mu_2 \cdots \mu_n} 
B^{\nu \mu_2 \cdots \mu_n} 
+ {\gamma \over n!a} R_{\mn\tau\sigma} 
B^{*\mn}{}_{\mu_3 \cdots \mu_n} 
B^{\tau\sigma \mu_3 \cdots \mu_n} \biggr], 
\label{action2}
\end{eqnarray}
where $\alpha$, $\beta$ and $\gamma$ are arbitrary constant 
parameters. The parameter $\mu$ is chosen as 
\begin{equation}
\mu = m - \left[ d(d-1) \alpha + (d-1) \beta 
+ 2 \gamma \right] a 
\end{equation}
so that this action coincides with eq.\ (\ref{action}) for 
the AdS metric. The last three terms in eq.\ (\ref{action2}) 
containing the curvature tensors are generalizations of the 
well-known $R \phi^2$ term in the scalar field theory. 
These terms give improvement terms in the energy-momentum tensor. 
Note that we do not need to introduce the term $R_{\mu\tau\nu\sigma} 
B^{*\mn}{}_{\mu_3 \cdots \mu_n} B^{\tau\sigma \mu_3 \cdots \mu_n}$ 
since it is related to the last term in eq.\ (\ref{action2}) by 
the Bianchi identity of the Riemann tensor. 
{}From eq.\ (\ref{action2}) we obtain the energy-momentum tensor as 
\begin{eqnarray}
T_{\mn} 
\A = \A - {2 \over \sqrt{-g}} {\delta S' \over \delta g^{\mn}} \nonu
\A = \A {2 \over n!} \left[ nm + (d-1)\beta a 
+ 2\gamma a \right] (B_{(\mu}^* B_{\nu)}) \nonu
\A\A - {1 \over n!} \left[ m - 2(d-1) \alpha a \right] 
g_{\mn} (B^* B) \nonu
\A\A + {2\alpha \over n!a} \left( D_\mu D_\nu 
- g_{\mn} D^2 \right) (B^* B) \nonu
\A\A + {\beta \over n!a} \biggl[ 2 D_{\sigma} D_{(\mu} 
(B^*_{\nu)} B^\sigma) - D^2 (B_{(\mu}^* B_{\nu)}) 
- g_{\mn} D_{(\tau} D_{\sigma)} (B^{*\tau} B^{\sigma}) 
\biggr] \nonu
\A\A - {4\gamma \over n!a} D_{\tau} D_{\sigma} 
(B^*_{(\mu}{}^\tau B_{\nu)}{}^{\sigma}), 
\label{emtensor}
\end{eqnarray}
where $(B^* B) = B^*_{\mu_1 \mu_2 \cdots \mu_n} 
B^{\mu_1 \mu_2 \cdots \mu_n}$, $(B^*_\mu B_\nu) = 
B^*_{\mu \rho_2 \cdots \rho_n} B_\nu{}^{\rho_2 \cdots \rho_n}$, etc. 
There is no contribution from the kinetic term. 
\par
The energy is defined as follows \cite{AD}. 
The energy-momentum tensor (\ref{emtensor}) satisfies 
$D_\mu T^{\mu\nu} = 0$ for arbitrary $\alpha$, $\beta$ and 
$\gamma$ when the field equation is used. 
We can construct a conserved current 
\begin{equation}
\partial_{\mu} \left( \sqrt{-g} T^{\mu}{}_{\nu} \xi^{\nu} \right) = 0 
\label{continuityeq}
\end{equation}
for each Killing vector $\xi^{\mu}$ satisfying 
$D_{\mu} \xi_{\nu} + D_{\nu} \xi_{\mu} = 0$. 
The energy is the charge of this current for a timelike Killing 
vector $\xi^\mu = (1, 0, \cdots, 0)$ 
\begin{equation}
E = - \int d^{d-1} x \sqrt{-g} T^t{}_t, 
\label{energy}
\end{equation}
where the integral is over $(d-1)$-dimensional space. 
%
%
\newsection{Solutions of the field equation}
We shall obtain the general solution of the field equation 
(\ref{fieldeq}). It is more convenient to rewrite the field 
equation in another form. Applying $\partial_{\mu_1}$ to 
eq.\ (\ref{fieldeq}) we obtain a constraint 
\begin{equation}
D_{\mu_1} B^{\mu_1 \cdots \mu_n} = 0, 
\label{constraint}
\end{equation}
where $D_{\mu}$ is the covariant derivative in AdS spacetime. 
Applying the differential operation in the first term of 
eq.\ (\ref{fieldeq}) to eq.\ (\ref{fieldeq}) again and 
using eq.\ (\ref{fieldeq}) in the second term we obtain 
the second order equation 
\begin{equation}
\left[ D_{\mu} D^{\mu} + n(n+1) a^2 - m^2 \right] 
B^{\mu_1 \cdots \mu_n} = 0. 
\label{fieldeq2}
\end{equation}
We first solve these equations and then substitute 
the solutions into the first order equation (\ref{fieldeq}) 
to obtain further conditions. 
\par
By the constraint (\ref{constraint}) the components 
$B_{ta_2 \cdots a_n}$ and $B_{t\rho a_3 \cdots a_n}$ are not 
independent but can be expressed by 
$B_{a_1 \cdots a_n}$ and $B_{\rho a_2 \cdots a_n}$. 
The field equation (\ref{fieldeq2}) for 
$[\mu_1 \cdots \mu_n] = [a_1 \cdots a_n]$, 
$[\rho a_2 \cdots a_n]$ gives 
\begin{eqnarray}
\A \A {\cal L}_1 B^{a_1 \cdots a_n} 
- {2(-1)^n n \over \sin^3\rho \cos\rho} \nabla^{[a_1} 
B^{a_2 \cdots a_n]\rho} = 0, \nonu
\A \A {\cal L}_2 B^{\rho a_2 \cdots a_n} 
- {2 \over \tan\rho} \nabla_{a_1} B^{a_1 a_2 \cdots a_n} = 0, 
\label{l12}
\end{eqnarray}
where $\nabla_a$ is the covariant derivative on S${}^{d-2}$ 
using the Christoffel connection $\gamma_{bc}^{\;a}$ in 
eq.\ (\ref{cristoffel}). The differential operators 
${\cal L}_1$ and ${\cal L}_2$ are defined as 
\begin{eqnarray}
{\cal L}_1 \A = \A - \partial_t^2 + \partial_\rho^2 
+ {4n-1 \over \sin\rho \cos\rho} \partial_\rho 
+ {\nabla_a \nabla^a + 3n(n-1) \over \sin^2\rho}
+ {4n^2 - \left( {m \over a} \right)^2 \over \cos^2\rho}, \nonu
{\cal L}_2 \A = \A - \partial_t^2 + \partial_\rho^2 
+ \left( 4 \tan\rho + {4n-3 \over \sin\rho \cos\rho} 
\right) \partial_\rho 
+ {\nabla_a \nabla^a + 3n^2-7n+3 \over \sin^2\rho} \nonu
\A\A + {(2n+1)^2 - \left( {m \over a} \right)^2 \over \cos^2\rho} 
- 4. 
\end{eqnarray}
It can be shown that when eqs.\ (\ref{constraint}) and 
(\ref{l12}) are satisfied, then the remaining components 
of eq.\ (\ref{fieldeq2}) are automatically satisfied. 
\par
%
%
We decompose the antisymmetric tensor field into transverse 
and longitudinal modes by using spherical harmonics 
$Y^{(l)}_{a_1 \cdots a_n}(\theta)$ 
for antisymmetric tensor fields on S${}^{d-2}$. 
The spherical harmonics are transverse 
$\nabla^{a_1} Y^{(l)}_{a_1 \cdots a_n} = 0$ and are eigenfunctions 
of the Laplacian $\nabla_a \nabla^a$ on S${}^{d-2}$ 
with the eigenvalue $-[l(l+d-3)-n]$. 
The quantum number $l$ takes values $l = 0, 1, 2, \cdots$ for $n=0$ 
and $l = 1, 2, 3, \cdots$ for $n \geq 1$. 
In the Appendix we sketch how to construct $Y^{(l)}_{a_1 \cdots a_n}$ 
and give some useful identities. 
Using the spherical harmonics the components of the antisymmetric 
tensor field are decomposed as 
\begin{eqnarray}
B^{a_1 \cdots a_n}(x) 
\A = \A R_1(t, \rho) Y^{(l) a_1 \cdots a_n}(\theta) 
+ R_2(t, \rho) \nabla^{[a_1} Y^{(l) a_2 \cdots a_n]}(\theta), \nonu
B^{\rho a_2 \cdots a_n}(x) 
\A = \A \sin \rho \cos \rho R_3(t, \rho) Y^{(l) a_2 \cdots a_n}(\theta) 
+ R_4(t, \rho) \nabla^{[a_2} Y^{(l) a_3 \cdots a_n]}(\theta), \nonu
B^{t a_2 \cdots a_n}(x) 
\A = \A R_5(t, \rho) Y^{(l) a_2 \cdots a_n}(\theta) 
+ R_6(t, \rho) \nabla^{[a_2} Y^{(l) a_3 \cdots a_n]}(\theta), \nonu
B^{t\rho a_3 \cdots a_n}(x) 
\A = \A R_7(t, \rho) Y^{(l) a_3 \cdots a_n}(\theta) 
+ R_8(t, \rho) \nabla^{[a_3} Y^{(l) a_4 \cdots a_n]}(\theta), 
\label{decomp}
\end{eqnarray}
where $Y^{(l) a_1 \cdots a_n} = h^{a_1b_1} \cdots h^{a_nb_n} 
Y^{(l)}_{b_1 \cdots b_n}$ and $\nabla^a = h^{ab} \nabla_b$. 
The factor $\sin\rho \cos\rho$ in front of $R_3$ is for later 
convenience. Substituting eq.\ (\ref{decomp}) into the constraint 
(\ref{constraint}) for $[\mu_2 \cdots \mu_n] = [a_2 \cdots a_n]$, 
$[\rho a_3 \cdots a_n]$, $R_5$, $\cdots$, $R_8$ are expressed 
in terms of $R_1$, $\cdots$, $R_4$ as 
\begin{eqnarray}
\partial_t R_5 \A = \A {1 \over n} (l+n-1)^2 R_2 
- \left( \sin\rho \cos\rho \partial_\rho + 2n \right) R_3, \nonu
\partial_t R_6 \A = \A - \left( 
\partial_\rho + 2 \tan\rho 
+ {2n-1 \over \sin\rho \cos\rho} \right) R_4, \nonu
\partial_t R_7 \A = \A - {1 \over n-1} 
(l+n)(l+n-2) R_4, \nonu
\partial_t R_8 \A = \A 0. 
\end{eqnarray}
Other components of the constraint (\ref{constraint}) are then 
automatically satisfied. 
$R_1$, $\cdots$, $R_4$, in turn, are determined by solving 
eq.\ (\ref{l12}). To solve eq.\ (\ref{l12}) it is convenient to 
change the variable as 
\begin{equation}
v = \sin^2 \rho. 
\end{equation}
\par
%
%
Let us first consider $R_1$. 
We define the function $f_1(v)$ by 
\begin{equation}
R_1(t, \rho) = \bar{N_1}(t) v^{{1 \over 2}\kappa} 
(1-v)^{{1 \over 2}\lambda} f_1(v), 
\end{equation}
where 
\begin{equation}
\bar{N_1}(t) = N_1 e^{-i\omega_1 t} + \tilde N_1 e^{i\omega_1 t} 
\end{equation}
and $N_1$ and $\tilde N_1$ are complex constants. 
The transverse part of the first equation in (\ref{l12}) gives 
\begin{eqnarray}
\A\A \biggl[ 4v(1-v) \partial_v^2 + 
2\left( 2\kappa +4n 
-2(\kappa + \lambda + 1) v \right) \partial_v 
+ \omega_1^2 - (\kappa+\lambda)^2 \nonu
\A\A {} \qquad + {(\kappa-l+n)(\kappa+l+3n-2) \over v} 
+ {(\lambda-2n)^2 
- \left( {m \over a} \right)^2 \over 1-v} \biggr] f_1(v) = 0. 
\label{r1eq}
\end{eqnarray}
We choose the parameters $\kappa$ and $\lambda$ as 
\begin{equation}
\kappa = l-n, \qquad
(\lambda-2n)^2 = \left( {m \over a} \right)^2 
\label{parameters}
\end{equation}
so that the $v^{-1}$ and $(1-v)^{-1}$ terms in eq.\ (\ref{r1eq}) 
vanish. There are two possible values of $\lambda$ 
\begin{equation}
\lambda = \lambda_{\pm} \equiv 2n \pm {|m| \over a}. 
\end{equation}
Then, eq.\ (\ref{r1eq}) becomes a hypergeometric equation 
\begin{equation}
\left[ v(1-v) \partial_v^2 + \left( c_1 - (a_1+b_1+1) v \right) 
\partial_v - a_1b_1 \right] f_1(v) = 0, 
\end{equation}
where 
\begin{eqnarray}
a_1 \A = \A {1 \over 2} ( \lambda + l - n - \omega_1 ), \nonu
b_1 \A = \A {1 \over 2} ( \lambda + l - n + \omega_1 ), \nonu
c_1 \A = \A l + n. 
\label{abc1}
\end{eqnarray}
The solution which gives $B_{a_1 \cdots a_n}$ regular at $\rho = 0$ 
is a hypergeometric function 
\begin{eqnarray}
f_1(v) \A = \A {}_2F_1 (a_1, b_1, c_1; v) \nonu
\A = \A {\Gamma(c_1) \over \Gamma(a_1) \Gamma(b_1)} 
\sum_{n=0}^\infty {\Gamma(a_1+n) \Gamma(b_1+n) 
\over \Gamma(c_1+n)} {v^n \over n!}. 
\label{hyperone}
\end{eqnarray}
\par
%
%
The equation for $R_4$ can be solved similarly. 
We put 
\begin{equation}
R_4(t, \rho) = \bar{N_4}(t) v^{{1 \over 2}(l-n+1)} 
(1-v)^{{1 \over 2}(\lambda+1)} f_4(v), 
\end{equation}
where 
\begin{equation}
\bar{N_4}(t) = N_4 e^{-i\omega_4 t} + \tilde N_4 e^{i\omega_4 t} 
\end{equation}
and $\lambda$ satisfies the second equation in (\ref{parameters}). 
The longitudinal part of the second equation in (\ref{l12}) gives 
a hypergeometric equation on $f_4$. The solution which gives 
$B_{\rho a_2 \cdots a_n}$ regular at $\rho = 0$ is 
\begin{equation}
f_4(v) = {}_2F_1(a_4, b_4, c_4; v), 
\label{hyperfour}
\end{equation}
where 
\begin{eqnarray}
a_4 \A = \A {1 \over 2} ( \lambda + l - n - \omega_4 ), \nonu
b_4 \A = \A {1 \over 2} ( \lambda + l - n + \omega_4 ), \nonu
c_4 \A = \A l + n. 
\label{abc4}
\end{eqnarray}
\par
%
%
The equations for $R_2$ and $R_3$ are slightly more complicated since 
they are coupled equations. The longitudinal part of the first equation 
and the transverse part of the second equation in (\ref{l12}) become 
\begin{eqnarray}
{\cal L} R_2 + {2n \over v} R_3 \A = \A 0, 
\nonu
{\cal L} R_3 + {2(l+n-1)^2 \over nv} R_2 \A = \A 0, 
\label{r23eq}
\end{eqnarray}
where 
\begin{eqnarray}
{\cal L} \A = \A 
-\partial_t^2 + 4v(1-v) \partial_v^2 
+ 4(2n-v) \partial_v \nonu 
\A \A 
- {l(l+2n-2) - (n-1)(3n+1) \over v} 
+ {4n^2 - \left( {m \over a} \right)^2 \over 1-v}. 
\end{eqnarray}
These equations can be diagonalized by defining new functions 
$\hat{R_2}$ and $\hat{R_3}$ as 
\begin{eqnarray}
\hat R_2 \A = \A -(l+n-1) R_2 + n R_3, \nonu
\hat R_3 \A = \A (l+n-1) R_2 + n R_3. 
\end{eqnarray}
Eq.\ (\ref{r23eq}) then becomes 
\begin{eqnarray}
\left[ {\cal L} - {2 \over v}(l+n-1) \right] {\hat R}_2 
\A = \A 0, \nonu 
\left[ {\cal L} + {2 \over v}(l+n-1) \right] {\hat R}_3 
\A = \A 0. 
\end{eqnarray}
Putting 
\begin{eqnarray}
\hat{R_2}(t, \rho) \A = \A \bar{N_2}(t) v^{{1 \over 2}(l-n+1)} 
(1-v)^{{1 \over 2}\lambda} f_2(v), \nonu
\hat{R_3}(t, \rho) \A = \A \bar{N_3}(t) v^{{1 \over 2}(l-n-1)} 
(1-v)^{{1 \over 2}\lambda} f_3(v), 
\end{eqnarray}
where 
\begin{eqnarray}
\bar{N_2}(t) \A = \A N_2 e^{-i\omega_2 t} 
+ \tilde N_2 e^{i\omega_2 t} , \nonu
\bar{N_3}(t) \A = \A N_3 e^{-i\omega_3 t} 
+ \tilde N_3 e^{i\omega_3 t}, 
\end{eqnarray}
these equations become hypergeometric equations on $f_2$ and $f_3$. 
The solutions which give $B_{a_1 \cdots a_n}$ and 
$B_{\rho a_2 \cdots a_n}$ regular at $\rho = 0$ are 
\begin{eqnarray}
f_2(v) \A = \A {}_2F_1 (a_2, b_2, c_2; v), \nonu
f_3(v) \A = \A {}_2F_1 (a_3, b_3, c_3; v), 
\label{hypertwothree}
\end{eqnarray}
where 
\begin{eqnarray}
a_2 \A = \A {1 \over 2} ( \lambda + l - n + 1 - \omega_2 ), \nonu
b_2 \A = \A {1 \over 2} ( \lambda + l - n + 1 + \omega_2 ), \nonu
c_2 \A = \A l+n+1, \nonu
a_3 \A = \A {1 \over 2} ( \lambda + l - n - 1 - \omega_3 ), \nonu
b_3 \A = \A {1 \over 2} ( \lambda + l - n - 1 + \omega_3 ), \nonu
c_3 \A = \A l+n-1. 
\label{abc23}
\end{eqnarray}
\par
Thus we have obtained the general solution of the second order 
equation (\ref{fieldeq2}). We now consider the first order 
equation (\ref{fieldeq}). Substituting eq.\ (\ref{decomp}) 
into eq.\ (\ref{fieldeq}) and using eq.\ (\ref{tdual}) and the second 
order equation (\ref{fieldeq2}) we find that eq.\ (\ref{fieldeq}) 
is satisfied if the functions $R$'s satisfy 
\begin{eqnarray}
{m \over a} R_4 
\A = \A (-1)^{{1 \over 2}n(n-1)} {n-1 \over l+n} 
\sin\rho \cos\rho \, \partial_t R_1, \nonu
{m \over a} R_3 
\A=\A (-1)^{{1 \over 2}n(n+1)+{1 \over 2}} {l+n-1 \over n} 
\left( \partial_t R_2 + {n \over \sin^2\rho} R_5 \right). 
\end{eqnarray}
These equations are satisfied if $\omega_1 = \omega_4$, 
$\omega_2 = \omega_3$ and $\bar N_i$ satisfy 
\begin{eqnarray}
{d \over dt} \bar N_1 \A = \A (-1)^{{1 \over 2}n(n-1)} 
{l+n \over n-1} {m \over a} \bar N_4, \nonu
{d \over dt} \bar N_3 \A = \A (-1)^{{1 \over 2}n(n+1)+{1 \over 2}} 
{a \over m} \biggl[ \, {1 \over 2} \biggl( (l+n-1)^2 
- \biggl( {m \over a} \biggr)^2 - \omega_2^2 \biggr) 
\bar N_3 \nonu
\A\A -2(l+n)(l+n-1) \bar N_2 \biggr]. 
\end{eqnarray}
By these relations $\bar N_4$ and $\bar N_3$ are related to 
$\bar N_1$ and $\bar N_2$ respectively and independent degrees 
of freedom are 
reduced from four to two. 
\par
%
%
\newsection{Conservation of the energy} 
We first consider the conservation of the energy as a condition 
for the stability. From eq.\ (\ref{continuityeq}) the time 
derivative of the energy (\ref{energy}) is given by 
\begin{equation}
{d \over dt} E = \int d^{d-2} \theta \left. \sqrt{-g} 
g^{\rho\rho} T_{\rho t} \right|_{\rho={\pi \over 2}}, 
\label{conservation}
\end{equation}
where the integral is over S${}^{d-2}$. 
$T_{\rho t}$ can be written as 
\begin{eqnarray}
n! T_{\rho t} \A = \A 2nm ( B_{(\rho}^* B_{t)} ) 
+ {2 \alpha \over a} \left( \partial_\rho - \tan\rho \right) 
\partial_t ( B^* B) \nonu
\A\A - {\beta \over a \sin\rho \cos\rho} \partial_t 
\left( B^*_{a_1 \cdots a_n} B^{a_1 \cdots a_n} \right) \nonu
\A\A + {\beta \over a} \left( \partial_\rho - \tan\rho 
- {n-1 \over \sin\rho \cos\rho} \right) 
\partial_t \left( B^*_{t a_2 \cdots a_n} 
B^{t a_2 \cdots a_n} \right) \nonu
\A\A + {\beta \over a} \left( \partial_\rho - \tan\rho 
+ {n \over \sin\rho \cos\rho} \right) 
\partial_t \left( B^*_{\rho a_2 \cdots a_n} 
B^{\rho a_2 \cdots a_n} \right) \nonu
\A\A + 2 (n-1) {\beta \over a} \left( \partial_\rho - \tan\rho 
+ {n+1 \over 2 \sin\rho \cos\rho} \right) 
\partial_t ( B^*_{t\rho} B^{t\rho} ) \nonu
\A\A - {4\gamma \over a \sin\rho\cos\rho} \partial_t \left( 
B^*_{t a_2 \cdots a_n} B^{t a_2 \cdots a_n} \right) 
+ 4 \gamma a ( B^*_{(\rho} B_{t)} ) \nonu
\A\A + {4 \gamma \over a} \left( \partial_\rho - \tan\rho 
+ {n+1 \over \sin\rho \cos\rho} \right) 
\partial_t ( B^*_{t\rho} B^{t\rho} ) + \nabla_a (\cdots)^a. 
\end{eqnarray}
Here, $\nabla_a (\cdots)^a$ represents total derivative terms 
on S${}^{d-2}$, which vanish in the integral 
(\ref{conservation}). 
Substituting eq.\ (\ref{decomp}) into eq.\ (\ref{conservation}) 
it is divided into three independent parts 
\begin{equation}
{d \over dt} E = {1 \over n!} \left. 
\int d\Omega \left[ 
\dot{E_1} \left| Y^{(l)}_{a_1 \cdots a_n} \right|^2 
+ \dot{E_2} \left| Y^{(l)}_{a_2 \cdots a_n} \right|^2 
+ \dot{E_3} \left| Y^{(l)}_{a_3 \cdots a_n} \right|^2 
\right] \right|_{\rho={\pi \over 2}}, 
\end{equation}
where $d\Omega = d^{d-2} \theta \sqrt{h}$ is the volume element 
of S${}^{d-2}$ and 
\begin{equation}
\left| Y^{(l)}_{a_1 \cdots a_n} \right|^2 = 
h^{a_1b_1} \cdots h^{a_nb_n} 
Y^{(l)*}_{a_1 \cdots a_n} Y^{(l)}_{b_1 \cdots b_n}.  
\end{equation}
$\dot E_1$, $\dot E_2$ and $\dot E_3$ depend only on 
$R_1$, ($R_2$, $R_3$) and $R_4$ respectively. 
\par
To evaluate the right hand side of eq.\ (\ref{conservation}) 
we need boundary behaviors of the functions $R_i$ for 
$\rho \rightarrow {\pi \over 2}$. They can be obtained from the 
behavior of the hypergeometric function ${}_2F_1(a,b,c;v)$ for 
$v \rightarrow 1$. When $\lambda-2n$ is not an integer,  
we find that near the boundary $R$'s behave as 
\begin{eqnarray}
R_1(\rho,t) \A \sim \A \bar{N_1}(t) \Bigl[ 
A_1 (\cos \rho)^{\lambda} \left( 1 + {\cal O}(\cos^2\rho)\right)
+ B_1 (\cos \rho)^{-\lambda+4n} 
\left( 1 + {\cal O}(\cos^2\rho)\right) \Bigr], \nonu
\hat R_2(\rho,t) \A \sim \A \bar{N_2}(t) \Bigl[ 
A_2 (\cos \rho)^{\lambda} 
\left( 1 + {\cal O}(\cos^2\rho)\right)
+ B_2 (\cos \rho)^{-\lambda+4n} 
\left( 1 + {\cal O}(\cos^2\rho)\right) \Bigr], \nonu
\hat R_3(\rho,t) \A \sim \A \bar{N_3}(t) \Bigl[ 
A_3 (\cos \rho)^{\lambda} 
\left( 1 + {\cal O}(\cos^2\rho)\right) 
+ B_3 (\cos \rho)^{-\lambda+4n} 
\left( 1 + {\cal O}(\cos^2\rho)\right) \Bigr], \nonu
R_4(\rho,t) \A \sim \A \bar{N_4}(t) \Bigl[ 
A_4 (\cos \rho)^{\lambda+1} 
\left( 1 + {\cal O}(\cos^2\rho)\right) 
+ B_4 (\cos \rho)^{-\lambda+4n+1} 
\left( 1 + {\cal O}(\cos^2\rho)\right) \Bigr], \nonu
\label{bbehavior}
\end{eqnarray}
where 
\begin{eqnarray}
\A \A A_i = {\Gamma(c_i)\Gamma(c_i-a_i-b_i) \over 
\Gamma(c_i-a_i)\Gamma(c_i-b_i)}, 
\ \ \ B_i = {\Gamma(c_i)\Gamma(a_i+b_i-c_i) \over 
\Gamma(a_i)\Gamma(b_i)}
\label{abcoeff} 
\end{eqnarray}
with $a_i$, $b_i$ and $c_i$ given in eqs.\ (\ref{abc1}), 
(\ref{abc4}) and (\ref{abc23}). We see that the value of 
$\lambda$ determines boundary behaviors of the solutions. 
The case in which $\lambda-2n$ 
is an integer is discussed at the end of this section. 
\par
Let us first consider $\dot{E_1}$. We obtain 
\begin{eqnarray}
\dot{E_1} 
\A = \A \left( {\tan\rho \over a}\right)^{4n-1} \left[ 
{2\alpha \over a} \left( \partial_\rho - \tan\rho \right) 
+ {4n\alpha - \beta \over a \sin\rho\cos\rho} \right] 
\partial_t |R_1|^2 \nonu
\A = \A {1 \over a^{4n}} \partial_t \left|\bar{N_1}\right|^2 
\Biggl[ - A_1^2  \biggl( 2(2\lambda-2n+1)\alpha+\beta \biggr) 
(\cos\rho)^{2\lambda-4n}\nonu
\A \A - 2 A_1B_1 \biggl( 2(2n+1)\alpha + \beta \biggr) \nonu
\A \A + B_1^2 \biggl( 2(2\lambda-6n-1)\alpha - \beta \biggr) 
(\cos\rho)^{-2\lambda+4n} \nonu
\A \A + C (\cos \rho)^{2\lambda -4n+2} 
+ D (\cos \rho)^2 + E (\cos \rho)^{-2\lambda +4n+2} 
\Biggr] \Biggr|_{\rho={\pi \over 2}}. 
\label{E1dot}
\end{eqnarray}
The last three terms represent higher order terms than the first 
three respectively. 
When we choose $\lambda = \lambda_+$, the first term automatically 
vanishes. For the second and the third terms to vanish 
we have to require either (i) $B_1 = 0$ or 
(ii) $A_1 = 0$, $2(2\lambda-6n-1)\alpha-\beta=0$. 
On the other hand, when we choose $\lambda = \lambda_-$, 
the third term vanishes automatically and we have to require 
either (iii) $A_1 = 0$ or 
(iv) $B_1 = 0$, $2(2\lambda-2n+1)\alpha+\beta=0$. 
It can be shown that the conditions (i) and (ii) are equivalent 
to (iii) and (iv) respectively. 
It is enough to consider two cases (i) and (iv) and set $B_1=0$. 
For $B_1=0$ only $C$ term survives among the higher order terms. 
In the case (i) it vanishes automatically. 
In the case (iv) we further need to require 
$2\lambda_- - 4n + 2 > 0$, i.e., $|m| < a$. 
\par
The conditions for $\dot{E_2}$ and $\dot{E_3}$ to vanish can be 
obtained in a similar way. It first requires $B_2 = B_3 = B_4 = 0$. 
The remaining terms are 
\begin{eqnarray}
\dot{E_2} 
\A = \A -{1 \over 2n a^{4n}} 
\left[ 2(2\lambda-2n+1)\alpha+\beta \right] 
(\cos\rho)^{2\lambda-4n} \partial_t 
| {\bar N_2} A_2 - {\bar N_3} A_3 |^2 \nonu
\A\A + {1 \over 8n^2 \omega_2^4 a^{4n}} \left[ 
2n(2\lambda-2n+1)\alpha + (2\lambda-n)\beta + 4\gamma \right] 
(\cos\rho)^{2\lambda-4n} \nonu
\A\A \times \partial_t | (\lambda-l-3n+1) {\dot{\bar N}}_2 A_2 
+ (\lambda+l-n-1) {\dot{\bar N}}_3 A_3 |^2 \nonu
\A\A + {1 \over 4n^2 \omega_2^2 a^{4n}} \left( 
2\gamma + {nm \over a} \right) 
\left( {\bar N_2}A_2 + {\bar N_3}A_3 \right)^* \nonu
\A\A \times \left[ (\lambda-l-3n+1) {\dot{\bar N}}_2 A_2 
+ (\lambda+l-n-1) {\dot{\bar N}}_3 A_3 \right] + {\rm c.c.}, \nonu
\dot{E_3} \A = \A -{1 \over \omega_1^2 a^{4n}} 
\partial_t \left|\bar{N_4}\right|^2 A_4^2 
(\cos \rho)^{2\lambda -4n} 
{1 \over n-1} (l+n)(l+n-2)(\lambda-2n) \nonu
\A \A \times \biggl[ 2n(2\lambda-2n+1)(\lambda-2n) \alpha + 
(2\lambda-n)(\lambda-2n)\beta \nonu
\A \A + \left( 4(\lambda-2n)-2 \right) \gamma 
- {nm \over a} \biggr] \biggr|_{\rho={\pi \over 2}}. 
\end{eqnarray}
For $\lambda = \lambda_+$ all these terms vanish automatically. 
For $\lambda = \lambda_-$ we have to require the coefficients 
to vanish, which fixes the parameters $\alpha$, $\beta$ 
and $\gamma$. 
\par
To summarize, the conservation of the energy leads to one 
of the two possibilities 
\begin{eqnarray}
{\rm (I)} \A\A \lambda = \lambda_+, \quad B_i = 0, \nonu
{\rm (II)} \A\A \lambda = \lambda_-, \quad B_i = 0, 
\quad  |m| < a. 
\label{result}
\end{eqnarray}
In the case (II) the coefficients of the improvement terms 
must be chosen as 
\begin{eqnarray}
\alpha \A = \A -{n \over 2(\lambda-n)(2\lambda-2n+1)} 
{m \over a}, \nonu
\beta \A = \A {n \over \lambda-n}{m \over a}, \nonu
\gamma \A = \A -{n \over 2}{m \over a}. 
\label{abg}
\end{eqnarray}
The conditions $B_i = 0$ require $a_i = 0$ or $b_i = 0$, 
which lead to the quantization of $\omega_i$ 
\begin{eqnarray}
\omega_1 \A = \A \pm (2k_1 + \lambda +l-n), \nonu
\omega_2 \A = \A \pm (2k_2 + \lambda +l-n + 1), \nonu
\omega_3 \A = \A \pm (2k_3 + \lambda +l-n - 1), \nonu
\omega_4 \A = \A \pm (2k_4 + \lambda +l-n), 
\label{discreteomega}
\end{eqnarray}
where $k_i$ are non-negative integers. 
Then, the  hypergeometric functions in eqs.\ (\ref{hyperone}), 
(\ref{hyperfour}), (\ref{hypertwothree}) can be expressed by the 
Jacobi polynomials as in the scalar field theory \cite{MT}. 
\par
The above analysis does not immediately apply to the 
case $\lambda-2n = N$ for an integer $N$. 
Let us consider the $N > 0$ case. 
(The $N < 0$ case is equivalent to the $N > 0$ case.) 
This occurs only when we choose $\lambda = \lambda_+$. 
In such a case the coefficients $A_i$ in eq.\ (\ref{abcoeff}) 
are divergent for generic values of $\omega_i$ since 
$c_i-a_i-b_i = -N$. To make $A_i$ finite we have to choose 
$\omega_i$ such that $c_i-a_i = -k'_i$ or $c_i-b_i = - k'_i$ 
for non-negative integers $k'_i$. 
The conservation of the energy requires $B_i=0$ as above, 
which restrict the values of $k'_i$ to $k'_i = N, N+1, \cdots$. 
Redefining $k_i = k'_i - N = 0, 1, \cdots$ we recover 
the values of $\omega_i$ in eq.\ (\ref{discreteomega}). 
Therefore, the results obtained for non-integer $\lambda-2n$ 
is also valid for integer $\lambda-2n$. 
\par
%
%
\newsection{Positivity of the energy} 
We next consider the second condition of stability, i.e., 
the positivity of the energy. The integrand of the energy 
(\ref{energy}) is 
\begin{eqnarray}
- \sqrt{-g} T^t{}_t 
\A = \A 2 m \sqrt{h} \left({\tan\rho \over a}\right)^{2n-1} 
\left[ n (B_t^* B_t) - {1 \over 2} g_{tt} (B^* B) \right] \nonu
\A\A + \partial_\rho \biggl[ \sqrt{h}  
\left({\tan\rho \over a}\right)^{2n-1} {1 \over a} \biggl\{ 
\left( \partial_\rho - \tan\rho \right) \left[ 
2\alpha (B^* B) + \beta (B^*_t B^t) \right] \nonu
\A\A + \left( \partial_\rho - \tan\rho 
+ {2n-1 \over \sin\rho \cos\rho} \right) \left[ 
\beta (B_\rho^* B^\rho) + 4\gamma (B^*_{t\rho} B^{t\rho}) 
\right] \nonu
\A\A - {1 \over \sin\rho \cos\rho} \left[ 
\beta (B_a^* B^a) + 4\gamma (B^*_{ta} B^{ta}) \right] 
\biggr\} \biggr] 
+ \sqrt{h} \nabla_a (\cdots)^a. 
\label{ttt}
\end{eqnarray}
We see that the contributions from the $\alpha$, $\beta$ and 
$\gamma$ terms in the action (\ref{action2}) are total 
derivative. Substituting eq.\ (\ref{decomp}) into 
eq.\ (\ref{ttt}) it is divided into three independent parts 
\begin{equation}
E = {1 \over n!} \int d\Omega \left[ 
E_1 \left| Y^{(l)}_{a_1 \cdots a_n} \right|^2 
+ E_2 \left| Y^{(l)}_{a_2 \cdots a_n} \right|^2 
+ E_3 \left| Y^{(l)}_{a_3 \cdots a_n} \right|^2 \right]. 
\end{equation}
$E_1$, $E_2$ and $E_3$ depend only on $R_1$, ($R_2$, $R_3$) 
and $R_4$ respectively. 
\par
$E_1$ is evaluated as 
\begin{equation}
E_1 
= m \int d\rho \left( {\tan \rho \over a} \right)^{4n+1} 
{1 \over \sin^2 \rho} \left|R_1\right|^2 
+ \Delta E_1, 
\label{eone}
\end{equation}
where $\Delta E_1$ is total derivative terms 
\begin{eqnarray}
\Delta E_1 
\A = \A \int d\rho \partial_\rho \left[ 
\left( {\tan \rho \over a} \right)^{4n-1} {1 \over a} 
\left\{ 2\alpha \left( \partial_\rho - \tan\rho 
+ {2n \over \sin\rho \cos\rho} \right) 
- {\beta \over \sin\rho \cos\rho} \right\} | R_1 |^2 
\right] \nonu
\A = \A \biggl[ - {1 \over a^{4n}} \left[ 
2(2\lambda-2n+1) \alpha + \beta \right] 
\left| \bar N_1 \right|^2 A_1^2 
(\cos \rho)^{2\lambda-4n} 
+ {\cal O}((\cos \rho)^{2\lambda-4n+2}) 
\biggr] \biggr|_{\rho={\pi \over 2}}. \nonu
\end{eqnarray}
The integral of the bulk term in eq.\ (\ref{eone}) is 
convergent for $\lambda = \lambda_+$ as seen from the 
boundary behavior of $R_1$ in eq.\ (\ref{bbehavior}). 
However, it is divergent for $\lambda = \lambda_-$. 
Therefore, the choice $\lambda = \lambda_-$ is not allowed 
and only $\lambda = \lambda_+$ is possible. 
In the case of scalar fields discussed in refs.\ \cite{BF,MT} 
the mass term in the energy is also divergent but it is 
canceled by a divergent contribution from the kinetic term. 
Both of $\lambda = \lambda_+$ and $\lambda = \lambda_-$ are 
possible in the scalar field theories. In the present theory 
there is no kinetic term in the energy since the kinetic 
term of the action is a topological term. 
For $\lambda = \lambda_+$ the bulk term in eq.\ (\ref{eone}) 
is obviously positive definite when $m > 0$. 
The boundary term $\Delta E_1$ vanishes since it has a 
positive power of $\cos\rho$. 
\par
Similarly, $E_2$ and $E_3$ are given by 
\begin{eqnarray}
E_2 \A = \A m \int d\rho 
\left( {\tan \rho \over a} \right)^{4n+1} 
{1 \over \sin^2 \rho} \biggl[ 
{1 \over n} (l+n-1)^2 \left|R_2\right|^2 \nonu
\A\A + n\left( \cos^2 \rho \left|R_3\right|^2 
+ {1 \over \sin^2 \rho} \left|R_5\right|^2 \right) \biggr] 
+ \Delta E_2, \nonu
E_3 \A = \A m \int d\rho 
\left( {\tan \rho \over a} \right)^{4n+1} 
{1 \over \sin^4\rho} 
\biggl[ n(n-1) {1 \over \sin^2 \rho} \left|R_7\right|^2 \nonu
\A\A + {n \over n-1} (l+n)(l+n-2) 
\left( \left|R_4\right|^2 + \left|R_6\right|^2 \right) 
\biggr] + \Delta E_3, 
\label{etwothree}
\end{eqnarray}
where the boundary terms are 
\begin{eqnarray}
\Delta E_2 
\A = \A -{1 \over 4na^{4n}} \biggl[ 
\left[ 2(2\lambda-2n+1)\alpha + \beta \right] 
\left| \bar N_2 A_2 - \bar N_3 A_3 \right|^2 
(\cos\rho)^{2\lambda-4n} \nonu
\A\A - {1 \over n\omega_2^4} \left[ 
2n(2\lambda-2n+1)\alpha+(2\lambda-n)\beta+4\gamma 
\right] \nonu
\A\A \times \left| (\lambda-l-3n+1) \dot{\bar N}_2 A_2 
+ (\lambda+l-n-1) \dot{\bar N}_3 A_3 \right|^2 
(\cos\rho)^{2\lambda-4n} \nonu
\A\A + {\cal O}((\cos\rho)^{2\lambda-4n+2}) 
\biggr] \biggr|_{\rho={\pi \over 2}}, \nonu
\Delta E_3 
\A = \A \biggl[ {(l+n) (l+n-2) (\lambda-2n)^2 \over (n-1) 
\omega_4^4 a^{4n}} \left[ 2n(2\lambda-2n+1)\alpha 
+ (2\lambda-n)\beta + 4\gamma \right] \nonu
\A\A \times \left| \dot{\bar N_4} \right|^2 A_4^2 
(\cos \rho)^{2\lambda-4n} 
+ {\cal O}((\cos\rho)^{2\lambda-4n+2}) \biggr] 
\biggr|_{\rho={\pi \over 2}}. 
\end{eqnarray}
For $\lambda = \lambda_+$ the bulk integrals in 
eq.\ (\ref{etwothree}) are convergent and are positive definite 
when $m > 0$. The boundary terms $\Delta E_2$ and $\Delta E_3$ 
vanish in the same way as $\Delta E_1$. 
\par
It is thus proved that the energy is well-defined and 
positive definite for the case (I) in eq.\ (\ref{result}) 
if the additional condition $m > 0$ is satisfied. 
The case (II) in eq.\ (\ref{result}) is not allowed since the
energy is divergent. Therefore, there exists only one complete 
set of solutions corresponding to the case (I) for antisymmetric 
tensor fields with the Chern-Simons type action. 
\par
%
%
\newsection{Scalar product} 
Finally we consider a scalar product of the fields. 
The scalar product of two solutions $B_{1\mu_1 \cdots \mu_n}$ 
and $B_{2\mu_1 \cdots \mu_n}$ is defined as 
\begin{equation}
\left( B_1, B_2 \right) 
= \int d^{d-1} x \sqrt{-g} J^t 
\label{sprod}
\end{equation}
by using the time component of the conserved current 
\begin{equation}
J^\mu = - i \left( 
B^*_{1 \nu_1 \cdots \nu_n} F^{\mu \nu_1 \cdots \nu_n}_2 
- F^{* \mu \nu_1 \cdots \nu_n}_1 B_{2 \nu_1 \cdots \nu_n} 
\right). 
\end{equation}
The integral in eq.\ (\ref{sprod}) is convergent since 
$\sqrt{-g} J^t = {\cal O}\left((\cos\rho)^{2\lambda-4n+1}\right)$ 
for $\rho \rightarrow {\pi \over 2}$ 
and $2\lambda-4n+1 > -1$ for $\lambda = \lambda_+$ and $m > 0$. 
The scalar product (\ref{sprod}) is also conserved 
\begin{equation}
{d \over dt} \left( B_1, B_2 \right) 
= - \int d^{d-2} \theta \left. \sqrt{-g} J^\rho 
\right|_{\rho={\pi \over 2}} 
= 0 
\end{equation}
since $\sqrt{-g} J^\rho|_{\rho={\pi \over 2}} 
= {\cal O}((\cos\rho)^{2\lambda-4n+2})|_{\rho={\pi \over 2}} 
= 0$ for $\lambda = \lambda_+$ and $m > 0$. 
Therefore we have a well-defined conserved scalar product. 
\par
%

\bigskip\bigskip

This work is partially supported  by the Grant-in-Aid from the  
Ministry of Education, Culture, Sports, Science and Technology, 
Japan, Priority Area (\#707) ``Supersymmetry and Unified Theory 
of Elementary Particles''. 

%
\def\numberbysectiona{\@addtoreset{equation}{section}
\def\theequation{A.\arabic{equation}}}
\numberbysectiona
\vspace{7mm}
\pagebreak[3]
\setcounter{section}{1}
\setcounter{equation}{0}
\setcounter{subsection}{0}
\setcounter{footnote}{0}
\begin{center}
{\large{\bf Appendix: Spherical harmonics on S${}^{d-2}$}}
\end{center}
\nopagebreak
\medskip
\nopagebreak
\hspace{3mm}
In this Appendix we sketch how to construct spherical harmonics 
for antisymmetric tensor fields on S${}^{d-2}$ following 
the approach in refs.\ \cite{GM,GP,RO}. See also ref.\ \cite{vN}. 
We embed a ($d-2$)-dimensional unit sphere S${}^{d-2}$ 
in ($d-1$)-dimensional Euclidean space {\bf R}${}^{d-1}$ with the 
Cartesian coordinates $x^i$ ($i = 1, 2, \cdots, d-1$). 
The metric of {\bf R}${}^{d-1}$ is given by 
\begin{eqnarray}
ds^2 \A = \A \delta_{ij} dx^i dx^j \nonu
\A = \A dr^2 + r^2 h_{ab}(\theta) d\theta^a d\theta^b, 
\end{eqnarray}
where $r=\sqrt{\delta_{ij}x^ix^j}$ is a radial coordinate and 
$\theta^a$ ($a = 1, 2, \cdots, d-2$) are angular coordinates 
parametrizing the unit sphere S${}^{d-2}$ with the metric $h_{ab}$. 
\par
Let us consider antisymmetric tensors $T_{i_1 \cdots i_n}$ 
in ${\bf R}^{d-1}$ which satisfy 
\begin{equation}
n^{i_1} T_{i_1 \cdots i_n} = 0, \qquad
\partial^{i_1} T_{i_1 \cdots i_n} = 0, 
\label{cartesiancond}
\end{equation}
where $n^i(\theta) = r^{-1} x^i$ is a unit vector normal to 
the sphere. In the polar coordinates ($r$, $\theta^a$) these 
conditions become 
\begin{equation}
T_{r a_2 \cdots a_n} = 0, \qquad
\nabla^{a_1} T_{a_1 \cdots a_n} = 0, 
\end{equation}
where $\nabla_a$ is the covariant derivative on S${}^{d-2}$. 
Therefore, restricting to the unit sphere they represent 
transverse tensors on the sphere. 
The relation between $T_{i_1 \cdots i_n}$ and 
$T_{a_1 \cdots a_n}$ is 
\begin{equation}
T_{a_1 \cdots a_n} 
= r^n {\partial n^{i_1} \over \partial \theta^{a_1}} 
\cdots {\partial n^{i_n} \over \partial \theta^{a_n}} 
T_{i_1 \cdots i_n}. 
\label{cprelation}
\end{equation}
\par
In the Cartesian coordinates spherical harmonics for such 
transverse antisymmetric tensors for $n \geq 1$ are given by 
\begin{equation}
Y^{(l)}_{i_1 \cdots i_n}(\theta) 
= r^{-l} C_{[i_1 \cdots i_n j_1](j_2 \cdots j_l)} 
x^{j_1} x^{j_2} \cdots x^{j_l}. 
\label{cartesiany}
\end{equation}
Here, $C_{[i_1 \cdots i_n j_1](j_2 \cdots j_l)}$ is a constant 
coefficient, which is antisymmetric in $i_1, \cdots, i_n, j_1$ 
and symmetric in $j_2, \cdots, j_l$, and is traceless with respect 
to any pair of the indices. 
Spherical harmonics for $n = 0$ are given by 
\begin{equation}
Y^{(l)}(\theta) 
= r^{-l} C_{(j_1 \cdots j_l)} x^{j_1} \cdots x^{j_l}, 
\label{cartesiany2}
\end{equation}
where $C_{(j_1 \cdots j_l)}$ is symmetric in $j_1, \cdots, j_l$ 
and is traceless with respect to any pair of the indices. 
Note that $l$ takes values $l = 0, 1, 2, \cdots$ for $n=0$ 
and $l = 1, 2, 3, \cdots$ for $n \geq 1$. 
One can easily check that these tensors indeed satisfy the 
conditions (\ref{cartesiancond}). Applying the Laplacian 
$\Delta_{d-1} = \delta^{ij} \partial_i \partial_j$ 
in ${\bf R}^{d-1}$ we find 
\begin{equation}
\Delta_{d-1} Y^{(l)}_{i_1 \cdots i_n} 
= -{l(l+d-3) \over r^2} Y^{(l)}_{i_1 \cdots i_n}. 
\label{cartesianeq}
\end{equation}
On the other hand, in the polar coordinates we have 
\begin{eqnarray}
\Delta_{d-1} Y^{(l)}_{a_1 \cdots a_n}
\A = \A \left[ {1 \over r^2} \nabla^a \nabla_a 
+ \left( \partial_r + {d-n-2 \over r} \right) 
\left( \partial_r - {n \over r} \right) 
- {n \over r^2} \right] Y^{(l)}_{a_1 \cdots a_n} \nonu
\A = \A {1 \over r^2} \left( \nabla^a \nabla_a - n \right) 
Y^{(l)}_{a_1 \cdots a_n}. 
\label{polareq}
\end{eqnarray}
In the last line we have used the fact that 
$Y^{(l)}_{a_1 \cdots a_n}$, which is related to 
eqs.\ (\ref{cartesiany}), (\ref{cartesiany2}) by the 
relation (\ref{cprelation}), has $r$ dependence $r^n$. 
Comparing eqs.\ (\ref{cartesianeq}) and (\ref{polareq}) we 
obtain eigenvalues of the Laplacian on S${}^{d-2}$ 
\begin{equation}
\nabla^a \nabla_a Y^{(l)}_{a_1 \cdots a_n} = 
-[l(l+d-3)-n] Y^{(l)}_{a_1 \cdots a_n}. 
\label{eigeneq}
\end{equation}
Useful identities, which can be easily derived from 
eq.\ (\ref{eigeneq}) are 
\begin{eqnarray}
\nabla^a \nabla_a \partial_{[a_1} Y^{(l)}_{a_2 \cdots a_n]} 
\A = \A - [l(l+d-3) - d + n + 2] \partial_{[a_1} 
Y^{(l)}_{a_1 \cdots a_n]},  \nonu
\nabla^{a_1} \nabla_{[a_1} Y^{(l)}_{a_2 \cdots a_n]} 
\A = \A - {1 \over n} (l+n-1) (l+d-n-2) Y^{(l)}_{a_2 \cdots a_n}. 
\label{useful}
\end{eqnarray}
\par
There is a duality relation between $Y^{(l)}_{a_1 \cdots a_m}$ 
and $Y^{(l)}_{a_1 \cdots a_{d-m-3}}$, which we use in sect.\ 3. 
By appropriately choosing the coefficients $C$'s in 
eqs.\ (\ref{cartesiany}) and (\ref{cartesiany2}) the relation 
can be written as 
\begin{equation}
Y^{(l)a_1 \cdots a_m} 
= {(-1)^{{1\over 2}(m+1)(d-m-2)} \over (l+m)(d-m-3)!} 
{1 \over \sqrt{h}} \epsilon^{a_1 \cdots a_{2n+1}} 
\partial_{a_{m+1}} Y^{(l)}_{a_{m+2} \cdots a_{d-2}}. 
\label{tdual}
\end{equation}
One can easily check that both hand sides of this equation 
are transverse and have the same eigenvalue of 
$\nabla_a \nabla^a$. The normalization factor on the right 
hand side is determined by repeated applications of this relation. 
%
%
%
\newcommand{\NP}[1]{{\it Nucl.\ Phys.\ }{\bf #1}}
\newcommand{\PL}[1]{{\it Phys.\ Lett.\ }{\bf #1}}
\newcommand{\CMP}[1]{{\it Commun.\ Math.\ Phys.\ }{\bf #1}}
\newcommand{\MPL}[1]{{\it Mod.\ Phys.\ Lett.\ }{\bf #1}}
\newcommand{\IJMP}[1]{{\it Int.\ J. Mod.\ Phys.\ }{\bf #1}}
\newcommand{\PR}[1]{{\it Phys.\ Rev.\ }{\bf #1}}
\newcommand{\PRL}[1]{{\it Phys.\ Rev.\ Lett.\ }{\bf #1}}
\newcommand{\PTP}[1]{{\it Prog.\ Theor.\ Phys.\ }{\bf #1}}
\newcommand{\PTPS}[1]{{\it Prog.\ Theor.\ Phys.\ Suppl.\ }{\bf #1}}
\newcommand{\AP}[1]{{\it Ann.\ Phys.\ }{\bf #1}}
\newcommand{\ATMP}[1]{{\it Adv.\ Theor.\ Math.\ Phys.\ }{\bf #1}}
\end{document}